\documentclass[twoside]{article}
\usepackage{multicol,graphicx,amssymb,mathrsfs,amsmath,pifont,amscd,latexsym,amsthm,geometry,color,fancyhdr,upgreek}
\input vatola.sty \input cyracc.def
\input psfig.sty

\newcommand{\sanhao}{\fontsize{19.08pt}{3\baselineskip}\selectfont}
\newcommand{\xiaosihao}{\fontsize{12pt}{\baselineskip}\selectfont}
\newcommand{\wuhao}{\fontsize{10.5pt}{\baselineskip}\selectfont}
\newcommand{\xiaowuhao}{\fontsize{9.5pt}{\baselineskip}\selectfont}
\newcommand{\dawuhao}{\fontsize{10pt}{0.8\baselineskip}\selectfont}
\newcommand{\liuhao}{\fontsize{7.875pt}{\baselineskip}\selectfont}

\newcommand{\qihao}{\fontsize{7pt}{\baselineskip}\selectfont}
\newcommand{\bahao}{\fontsize{8pt}{\baselineskip}\selectfont}
\TagsOnRight

\renewcommand{\baselinestretch}{1.06} 
\catcode`@=11 \long\def\@makefntext#1{\noindent #1}
\newskip\tabcentering \tabcentering=1000pt plus 1000pt minus 1000pt
\def\REF#1{\par\hangindent\parindent\indent\llap{#1\enspace}}
\def\MCH#1#2{\setbox0=\hbox{\raise#1\hbox{#2}}\smash{\box0}}

\def\@evenfoot{}\def\@oddfoot{}
\def\@evenfoot{\vbox{\hbox to \textwidth{\bahao\sf\hbox to
0.01cm{\textbf{\thepage}\hfill} \hfill{\emph{Chen, P. F. Sci
China Ser G-Phys Mech Astron }{$|$ Nov. 2009 $|$ vol. 52 $|$ no. 11
$|$ \textbf{1785-1789}}}\hfill}}}
\def\@oddfoot{\vbox{\hbox to \textwidth{\bahao\sf\hbox to
0.01cm{} \hfill{ \emph{Chen, P. F. Sci China Ser G-Phys Mech
Astron} {$|$ Nov. 2009 $|$ vol. 52 $|$ no. 11 $|$
\textbf{1785-1789}}}\hfill\textbf{\thepage}}}}

\def\sec#1{\vspace{6mm}\noindent{{\xiaosihao\sf\textbf{#1}}}\vspace{2mm}}

  \def\tlj{\end{document}}  \newsymbol\wjzhml 203F

\def\tlj{\end{document}}
\def\no{\noindent}

\begin{document}
\abovedisplayskip=5pt plus 2pt minus 3pt 
\belowdisplayskip=5pt plus 2pt minus 3pt 
\columnsep 20pt
\textwidth=176truemm \textheight=229truemm

\renewcommand{\baselinestretch}{0.9}\baselineskip 9pt
{\psfig{figure=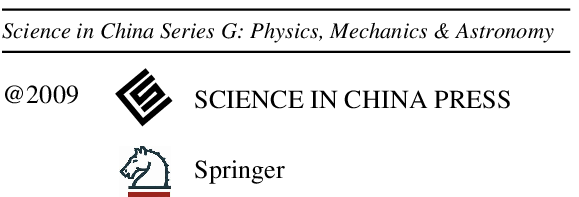}\hfill
\begin{picture}(43,0)
\rightline{\sf\put(-50,25){{\vbox{\hbox {\hspace{5.85mm}\dawuhao
www.scichina.com} \hbox{\hspace{6.5mm}\dawuhao phys.scichina.com}
\hbox{\,\dawuhao www.springerlink.com}}}}}
\end{picture}


\vspace{11.6true mm}
\renewcommand{\baselinestretch}{0.7}\baselineskip 5pt
\noindent{\parbox{16.3cm}{\sanhao{\sf\textbf{EIT waves and coronal magnetic field diagnostics}}}}
\vspace{0.5 true cm}
\renewcommand{\sfdefault}{phv}

\renewcommand{\baselinestretch}{1.6}\baselineskip 19.08pt
\noindent{\sf CHEN PengFei\footnotetext{ \baselineskip 6pt \qihao
\vspace{-2.2mm}\\
Received July 15, 2009; accepted August 15, 2009\\
doi: 10.1007/s11433-009-0240-9\\
$^\dag$Corresponding author (email: chenpf@nju.edu.cn)\\
Supported by the 973 project (Grant No. 2006CB806302), and the National Natural Science
Foundation of China (Grant Nos. 10933003 and 10403003)\vspace{2mm}\\
}

\renewcommand{\headrulewidth}{0pt}
\thispagestyle{fancy}\fancyfoot{} \fancyfoot[L]{\scriptsize
\vspace{-10mm}
\fcolorbox[rgb]{0,0,0}{0.85,0.85,0.85}{\parbox[t]{160truemm{\bf
Citation:\ }}{ \baselineskip 12pt
\renewcommand{\baselinestretch}{1.1} \scriptsize
Chen, P. F. EIT waves and coronal magnetic field diagnostics. Sci
China Ser G, 2009, 52(11): 1785-1789, doi: 10.1007/s11433-009-0240-9}}
}\rm

\vspace{0.2 true cm}\noindent
\parbox{16.3cm}
{\noindent\renewcommand{\baselinestretch}{1.3}\baselineskip 12pt
{\liuhao\sf
Department of Astronomy, Nanjing University, Nanjing \ 210093, China\vspace{2mm}}}

\noindent{\xiaowuhao\sf\textbf{\hspace{-1mm}
\parbox{16.3cm}
{\noindent
\renewcommand{\baselinestretch}{1.3}\baselineskip 13pt
Magnetic field in the solar lower atmosphere can be measured by the
use of the Zeeman and Hanle effects. In contrast, the coronal
magnetic field well above the solar surface, which directly controls
various eruptive phenomena, can not be precisely measureed with the
traditional techniques. Several attempts are being made to probe the
coronal magnetic field, such as force-free extrapolation based on
the photospheric magnetograms, gyroresonance radio emissions, and
coronal seismology based on MHD waves in the corona. Compared to the
waves trapped in the localized coronal loops, EIT waves are the only
global-scale wave phenomenon, and thus are the ideal tool for the
coronal global seismology. In this paper, we review the observations
and modelings of EIT waves, and illustrate how they can be applied
to probe the global magnetic field in the corona.}}}

\vspace{5.5mm}{\noindent \footnotesize \sf
\parbox{16.3cm}
{\noindent
\renewcommand{\baselinestretch}{1.3}\baselineskip 13pt
coronal mass ejections (CMEs), magnetic field, diagnostics,
magnetohydrodynamics (MHD)}} \vspace{6mm} \baselineskip 15pt

\begin{multicols}{2}

\renewcommand{\baselinestretch}{1.08}
\parindent=10.8pt  
\rm\wuhao  \vspace{-4mm}

\noindent Magnetic field plays a vital role in various astrophysical
processes. In the solar atmosphere, magnetic field interacts with
the plasma, producing abundant eruptive activities, besides the
relatively quiescent heating in the atmosphere. Therefore, the
measurement of magnetic field in the solar atmosphere is extremely
important. For strong magnetic field, Zeeman effect has long been
used in the solar photosphere and chromosphere, and is being
testified for the transition region. For weak magnetic field
measurement, Hanle effect can be utilized$^{[1]}$.  Spectral lines
are formed in a relatively narrow height range in these atmospheric
layers, so the local magnetic field can be obtained directly.
However, for the corona, where magnetic field directly controls the
plasma kinematics, optical emission lines are optically thin, and
the traditional
techniques do not work so well$^{[2]}$. Although
some techniques have been developed to extrapolate the coronal
magnetic field based on photospheric magnetograms, the extrapolation
is an ill-posed problem anyway. Therefore, it is imperative to
develop alternative approaches to probe the coronal magnetic field,
such as gyroresonance radio emissions.

Ubiquitous perturbations in the corona would propagate in the form
of MHD waves in the quiet Sun or trigger the oscillations of coronal
structures, such as prominences and coronal loops, caused by the
trapped MHD waves. The phase speeds of the propagating waves or the
standing waves are determined by the local magnetic field and/or
temperature. Thus, either the propagation speed or the oscillation
period provides effective information to diagnose the magnetic field
in the corona, which gives rise to a new discipline, i.e, coronal
seismology$^{[3]}$.

In the early 1960s, Moreton \& Ramsey$^{[4]}$ discovered a kind of
arc-shaped chromospheric perturbations propagating away from big
flares to a distance $>5\times 10^5$ km, later called Moreton waves.
According to Uchida$^{[5]}$, Moreton waves were due to coronal
fast-mode waves sweeping the chromosphere, and they can be used to
reconstruct the Alfv\'en velocity, hence the magnetic field
distributions in the corona. In contrast to this large-scale waves,
localized MHD waves were also found to be trapped in coronal loops
and prominences (see ref. [6] for a review).

Coronal loop/prominence oscillations provide the diagnosis of the
local magnetic field, whereas Moreton waves provide the magnetic
field information on a relatively larger scale. Both phenomena cover
only a small part of the solar surface. However, EIT waves, which
were discovered in the coronal mass ejection (CME) event on 1997 May
12, offer a perfect opportunity for the magnetic field diagnosis on
a global scale.

\sec{1\quad EIT waves: Observations and modelings}

\noindent After the launch of the SOHO satellite in late 1995, one
of its payloads, the EUV Imaging Telescope (EIT), detected an almost
circular front propagating outward from the source active region in
the CME/flare event on 1997 May 12$^{[7]}$. More frequently, they
are present as propagating patches, rather than circular fronts.
Such a wave phenomenon is often referred to as ``EIT wave", but
occasionally called ``coronal wave" by some authors (e.g., ref.
[8]). EIT wave fronts, with an emission enhancement ranging from a
few to tens of percent, are always followed immediately by expanding
EUV dimmings in the base difference images$^{[9]}$. Therefore, EIT
waves and dimmings are considered symbiotic phenomena which require
a common physical process in any theoretical model$^{[10]}$.

Since EIT waves and dimmings are found visible in several EUV bands,
they are believed to result mainly from density enhancement and
depletion, respectively, although Wills-Davey \& Thompson$^{[11]}$
and Chen \& Fang$^{[12]}$ pointed out that the temperature variation
also plays a role in determining the EUV intensity. The statistical
study by Klassen et al.$^{[13]}$, using SOHO/EIT observations,
indicates that the typical velocities of EIT waves range from 170 to
350 km$\cdot$s$^{-1}$, with a mean value of 271 km$\cdot$s$^{-1}$.
Note that the EIT wave velocity can be as slow as 50
km$\cdot$s$^{-1}$ or even smaller$^{[14]}$. According to recent
research by Long et al.$^{[15]}$, who analyzed the high-cadence EUV
observations by STEREO satellite, the low cadence of the SOHO/EIT
observations smoothed the EIT wave velocity evolution, i.e, the
maximum velocity was underestimated and the low velocity was
overestimated. Therefore, a renewed statistics of EIT wave
velocities, based on the high cadence observations of STEREO/SECCHI,
is strongly recommended.

The discovery of EIT waves immediately reminded the researchers of
the coronal counterparts of H$\alpha$ Moreton waves, and so they
were explained in terms of fast-mode MHD waves in the
corona$^{[9,16,17]}$, though there were arguments whether the
fast-mode waves originate from the flare or the CME$^{[18,19]}$.
However, the fast-mode wave model can not account for the following
observational features: (1) Even considering that the high-cadence
observations may upgrade the statistical results made by Klassen et
al.$^{[13]}$, the EIT wave velocities are still significantly
smaller than those of Moreton waves, which are truly fast-mode; (2)
EIT waves may stop at the footpoints of the magnetic
separatrix$^{[20]}$; (3) EIT wave velocities are anti-correlated
with the speeds of the corresponding type II radio bursts$^{[13]}$;
(4) EIT waves accelerate from $\sim20$ km$\cdot$s$^{-1}$ near the
source active region to more than 400 km$\cdot$s$^{-1}$ in the quiet
region$^{[15]}$.

In order to reconcile these discrepancies, Chen et al.$^{[21,22]}$
proposed a field-line stretching model, i.e,  as the strongly
twisted flux rope erupts to form a CME, a fast-mode piston-driven
shock propagates outward, which corresponds to the coronal
counterpart of H$\alpha$ Moreton wave. At the same time, the
erupting flux rope pushes the overlying magnetic loop successively.
The stretching of each magnetic loop leads to an EIT wave front on
the outer side and density depletion inside the magnetic loop.
Therefore, the model explains the EIT waves and dimmings
simultaneously. The field-line stretching model can account for all
the four features mentioned above, and was supported by imaging and
spectroscopic observations$^{[23]}$.

\sec{3\quad EIT waves as a tool for coronal seismology}

\noindent Since EIT waves propagate across almost the entire solar
disk in many events, they are the ideal tool for the global magnetic
field diagnosis. As for other wave phenomena, how EIT waves are used
for coronal seismology depends on the identification of the wave
mode. If they are fast-mode MHD waves, the magnetic field
reconstruction would be the same as used for Moreton waves. Within
such a framework, several groups tried to reconstruct the global
magnetic field distribution$^{[24]}$. Considering the discrepancies
between the fast-mode wave model and the observations, we have
proposed a magnetic field-line stretching model, as described in
Sec. 2. In this section, we discuss how EIT waves can be used for
coronal seismology in the framework of our field-line stretching
model.

The sketch of our model is depicted in Figure 1, where the thick
curved lines are the initial magnetic field lines. As the flux rope
({\it the circles}) moves upward, it pushes the top part of the
first field line near point $A,$ and this local part expands
outward. As shown in Figure 1(a), the expansion would be transferred
along the field line to its footpoints $C$ and $E,$ with the
Alfv\'enic velocity $v_{\rm A}$. The expansion near the footpoints
$C$ and $E$ compresses the plasma on the outer side, which produces
the first EIT wave front. At the same time, the local expansion near
point $A$ would also be transferred upward to the top part of the
second magnetic field line, near point $B,$ with the fast-mode wave
velocity ($v_{\rm f}=\sqrt{v_{\rm A}^2+v_{\rm s}^2}$, where $v_{\rm
s}$ is the sound speed) since it is perpendicular to the field
lines, as illustrated by the right panel. Thus the top part of the
second field line near point $B$ expands. Again, the local
expansion, as a kind of perturbation, would be transferred along the
second field line to its footpoints $D$ and $F,$ with the Alfv\'enic
velocity $v_A$. Thereby, a second EIT wave front is formed near
points $D$ and $F.$ With such a repeating circle, progressing EIT
wave fronts are formed as outer field lines are pushed to expand
successively. Note that the successive expansion of the magnetic
loops naturally leads to expanding dimming regions behind the EIT
wave fronts. The apparent velocity of the EIT wave in the model is
the distance between two successive EIT wave fronts, e.g., CD,
divided by the time difference between the formation of the two
successive EIT wave fronts, $\Delta t$. This is to say, the EIT wave
apparent velocity near points $C$ and $D$ is related to the magnetic
field by $v_{\rm EIT}=CD/\Delta t$, where $\Delta t=\int_A^B1/v_f
{\rm d}s+\int_B^D1/v_A {\rm d}s-\int_A^C1/v_A {\rm d}s$.

\end{multicols}

\vspace*{3mm}
\begin{center}
\centerline{\psfig{figure=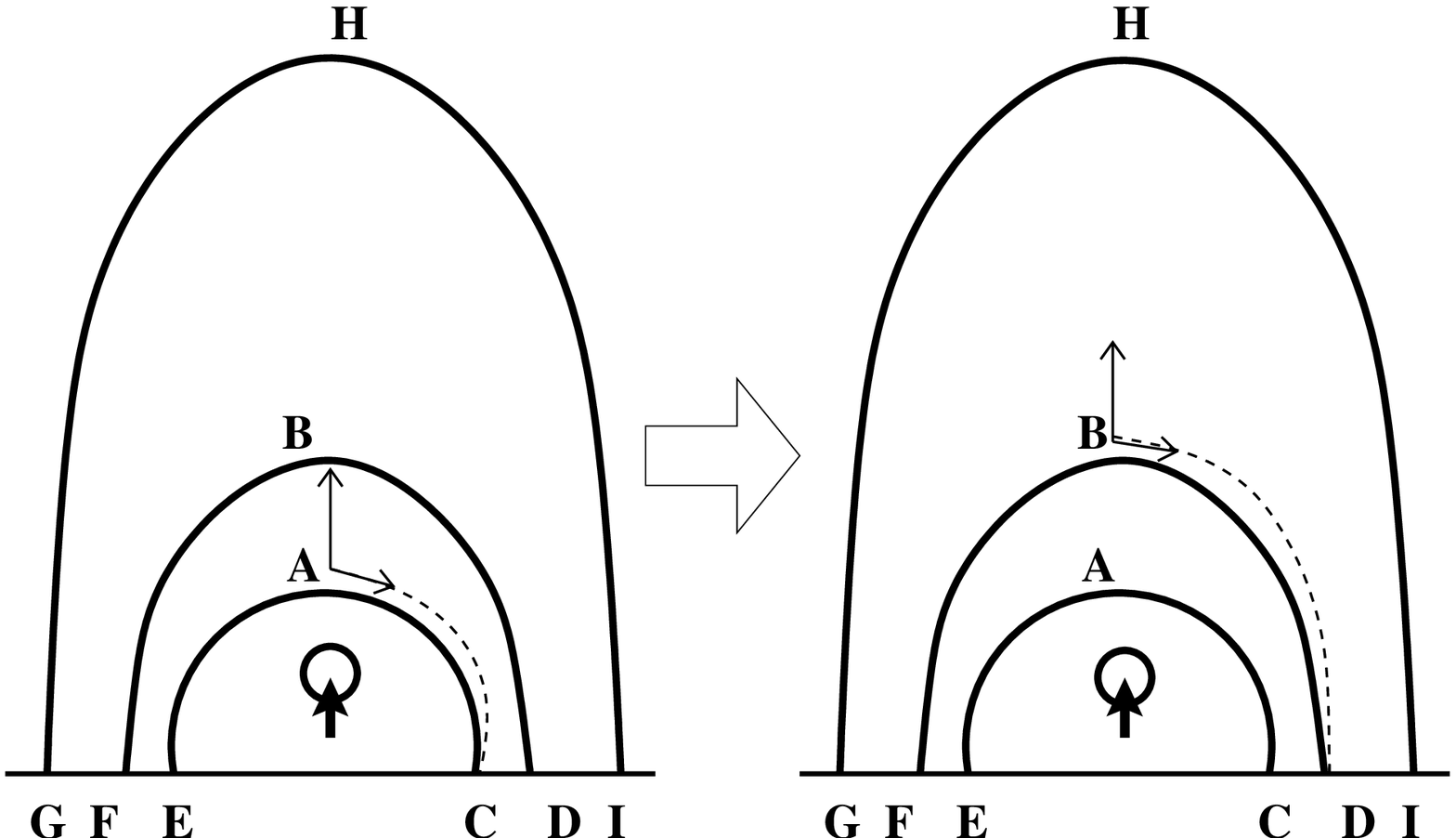,width=11 cm}}
\centerline{\footnotesize {\bf Figure 1}\quad The sketch of our
field-line stretching model for EIT waves$^{[21]}$.}
\end{center}

\begin{center}
\centerline{\psfig{figure=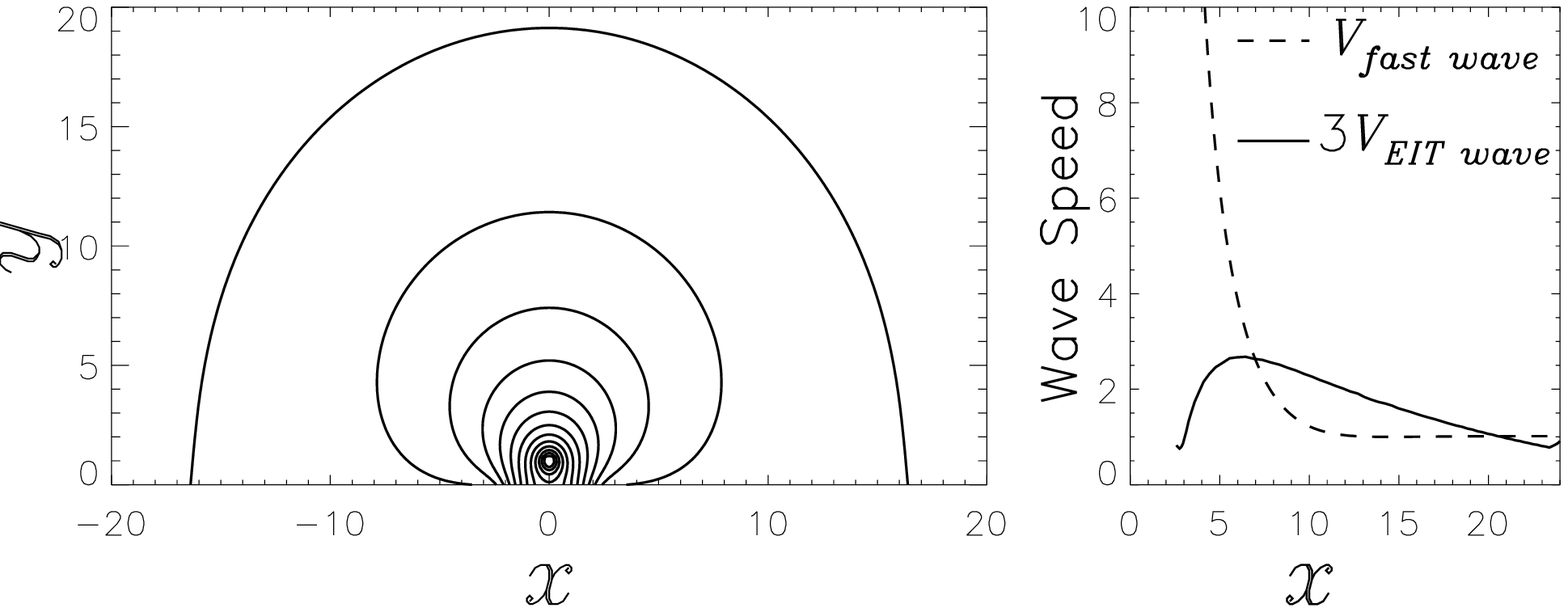,width=15 cm}}\end{center}\vspace*{-8mm}
{\footnotesize {\bf Figure 2}\quad (a) A magnetic configuration with
an active region surrounded by potential field; (b) the
corresponding distributions of the fast-mode wave and EIT wave
velocities along the surface. The unit of the velocity is 900
km$\cdot$s$^{-1}$.}

\begin{multicols}{2}

\renewcommand{\baselinestretch}{1.08}
\parindent=10.8pt  
\rm\wuhao  \vspace{-4mm}

As demonstrated by Chen et al.$^{[21]}$, if the magnetic field lines
are concentric semicircles, the resulting $v_{\rm EIT}$ is about
0.34$v_{\rm f}$. For any given magnetic configuration, the EIT wave
velocity distribution can be analogically derived by the
above\linebreak formula. For example, Figure 2(a) depicts
the\linebreak magnetic configuration for an~ active~ region~ sur-
rounded by a potential bipolar field.~ The corresponding EIT
wave velocity and the fast-mode wave velocity distributions are
plotted in the Figure 2(b). The EIT waves are seen to accelerate
near the boundary of the source active region. This feature is
consistent with the recent observation$^{[15]}$, but which cannot be
interpreted by the fast-mode wave model.

We can adjust the coronal magnetic configuration so that the
resulting EIT wave velocity distribution can best match the
observational one. In this way, the coronal magnetic field
distribution can be reconstructed. This procedure, which is forward
modeling, is demonstrated here to be eligible. The inverse modeling,
i.e., to derive the 3D coronal magnetic field directly from the
observed EIT wave velocity distribution, similar to the techniques
used in helioseismology, deserves substantial further studies.

\sec{4\quad Prospects}

\noindent As the largest-scale wave phenomenon in the corona, EIT
waves offer an ideal opportunity to probe the coronal magnetic
field. As proposed in our field-line stretching model, the formation
of each EIT wave front conveys the magnetic information in 3D space,
all the way from the erupting flux rope to the top of each field
line, and then along the field line to its footpoints. Therefore,
similar to the {\it p}-mode waves in the helioseismology, EIT waves
can be used to probe the 3D distribution of the coronal magnetic
field, providing that high-cadence EUV imaging observations are
available. In this paper, we demonstrated the feasibility of the
forward modeling of the coronal seismology based on EIT wave.
However, the inverse modeling, i.e., the direct inversion from EIT
wave velocity evolution to the coronal magnetic field, deserves
further investigation.

\end{multicols}

\begin{multicols}{2}
\normalsize \vskip0.16in\parskip=0mm \baselineskip 15pt
\renewcommand{\baselinestretch}{1.12}
 \footnotesize
\parindent=4mm

\bahao\REF{1\ }Sahal-Br\'echot S. The Hanle effect applied to
magnetic field diagnostics.  Space Sci Rev, 1981, 29: 391--401

\REF{2\ }White S M. Solar and Space Weather Radiophysics, Vol.314.
Gary D E, Keller C U, eds. Dordrecht: Academic Publisher, 2004.
89--113

\REF{3\ }Roberts B, Edwin P M, Benz A O. On coronal oscillations.
 Astrophys J, 1984, 279: 857--865

\REF{4\ }Moreton G E, Ramsey H E. Recent observations of dynamical
    phenomena associated with solar flares. Publ Astron Soc Pac, 1960, 72: 357

\REF{5\ }Uchida Y. Propagation of hydromagnetic disturbances in the
    solar corona and Moreton's wave phenomenon. Sol Phys, 1968,
    4: 30--44

\REF{6\ }Roberts B. Waves and oscillations in the corona. Sol Phys,
2000, 193: 139--152

\REF{7\ }Thompson B J, Plunkett S P, Gurman J B et al. SOHO/EIT
    observations of an Earth-directed coronal mass ejection on May 12,
    1997. Geophys Res Lett, 1998, 25: 2465--2468

\REF{8\ }Vr\v{s}nak B, Warmuth A, Temmer M et al. Multi-wavelength study
    of coronal waves associated with the CME-flare event of 3 November
    2003. Astron Astrophys, 2006, 448: 739--752

\REF{9\ }Thompson B J, Gurman J B, Neupert W M et al. SOHO/EIT
    observations of the 1997 April 7 coronal transient: Possible
    evidence of coronal moreton waves. Astrophys J, 1999, 517: 151--154

\REF{10\ }Chen P F. Initiation and propagation of CMEs. J Astrophys
Astron, 2008, 29: 179--186

\REF{11\ }Wills-Davey M J, Thompson B J. Observations of a
propagating disturbance in TRACE. Sol Phys, 1999, 190: 467--483

\REF{12\ }Chen P F, Fang C. EIT waves - A signature of global
magnetic    restructuring in CMEs. IAU Symp, 2005, 226: 55--64

\REF{13\ }Klassen A, Aurass H, Mann G et al. Catalogue of the 1997
    SOHO-EIT coronal transient waves and associated type II radio burst
    spectra. Astron Astrophys, 2000, 141: 357--369

\REF{14\ }Thompson B J, Myers D C. Catalog of Coronal ``EIT Wave''
    Transients. Astrophys J Suppl Ser, 2009, 183: 225--243

\REF{15\ }Long D M, Gallagher P T, McAteer R T J et al. The kinematics
    of a globally propagating disturbance in the solar corona. Astrophys
    J, 2008, 680: L81-L84

\REF{16\ }Wu S T, Zheng H, Wang S et al. 3D numerical simulation of MHD
   waves observed by EIT. J Geophys Res, 2001, 106: 25089-25102

\REF{17\ }Wang Y M. EIT waves and fast-mode propagation in the solar
    corona. Astrophys J, 2000, 543: L89--L92

\REF{18\ }Grechnev V V, Uralov A M, Slemzin V A et al. Absorption phenomena
    and a probable blast wave in the 13 July 2004 eruptive event. 
    Sol  Phys, 2008, 253: 263--290

\REF{19\ }Veronig A M, Temmer M, Vr\v{s}nak B. High-cadence
    observations of a global coronal Wave by STEREO EUVI.
   Astrophys J, 2008, 681: L113--L116

\REF{20\ }Delann\'ee C, Aulanier G. CME Associated with
    transequatorial loops and a bald patch flare. Sol Phys,
    1999, 190: 107--129

\REF{21\ }Chen P F, Wu S T, Shibata K, et al. Evidence of EIT and
    Moreton waves in numerical simulations. Astrophys J, 2002, 572: L99--L102

\REF{22\ }Chen P F, Fang C, Shibata K. A full view of EIT waves.
Astrophys J, 2005, 622: 1202--1210

\REF{23\ }Harra L K, Sterling A C. Imaging and spectroscopic
    investigations of a solar coronal wave. Astrophys J, 2003, 587: 429--438

\REF{24\ }Warmuth A, Mann G. A model of the Alfv\'en speed in the
    solar corona. Astron Astrophys, 2005, 435: 1123--1135
\end{multicols}
 \tlj